\documentclass[10pt,aps,superscriptaddress]{revtex4-2}
\usepackage{amssymb,amsmath}
\usepackage{graphicx}

\usepackage[urlcolor=purple,colorlinks=true,citecolor=blue,	linkcolor=red,pdfstartview={FitH},bookmarks=false]{hyperref}

\begin{document}

\title{Coordinate transformations and matter waves cloaking}

\author{G. R. Mohammadi}
\affiliation{Institute for Advanced Studies in Basic Sciences (IASBS), Zanjan 45137-66731, Iran}
\author{A. G. Moghaddam}
\affiliation{Institute for Advanced Studies in Basic Sciences (IASBS), Zanjan 45137-66731, Iran}
\author{R. Mohammadkhani}
\affiliation{Department of Physics, Faculty of Science, University of Zanjan, Zanjan 45371-38791, Iran}

\date{11 January 2016}

\begin{abstract}
Transformation method provides an efficient tool to control wave propagation inside the materials. Using the coordinate transformation approach, we study invisibility cloaks with sphere, cylinder and ellipsoid structures for electronic waves propagation. The underlying physics behind this investigation is the fact that Schr\"odinger equation with position dependent mass tensor and potentials have a covariant form which follows the coordinate transformation. Using this technique we obtain the exact spatial form of the mass tensor and potentials for a variety of cloaks with different shapes. 
\end{abstract}

\maketitle
\section{Introduction}
The theory of wave propagation and controlling them can be expressed in a unified framework irrespective of the nature of waves varying from optical, acoustic, and even matter waves. The basic idea of theoretical methods for cloaking and controlling waves is to use the covariant form of the corresponding wave equation and then employ a suitable space-time transformation. 
In reality we cannot deform the space to control wave propagation and the corresponding metrics cannot be changed. Alternatively, we can use the inverse of the obtained coordinate transformation to find the spatial or in some cases the temporal variations of the medium parameters in order to get the desired wave propagation. The medium parameters depending on the characteristics of the wave could be different. However in the most studied case of electromagnetic waves they are electrical permittivity and magnetic permeability \cite{ward1996refraction}. Considering the covariant form of wave equations which is satisfied in all frames, the transformation from the Cartesian coordinate system to generalized curvilinear coordinates can be easily obtained.
Then by applying the inverse transformation, the material constitutive parameters can be obtained for a cloaking in the real physical space.

Two decades ago Pendry's proposed a theoretical model in order to have an invisibility cloaks for the electromagnetic waves \cite{ward1996refraction}. Further studies showed how one can control electromagnetic fields in general situations by manipulating the material properties which mathematically mimics a general coordinate transformation \cite{pendry2006controlling,leonhardt2006general,schurig2006calculation,leonhardt2006optical,
leonhardt2006notes,schurig2007transformation}. The basic principle of the so-called transformation optics for electromagnetic wave propagation is the fact that Maxwell's equations have a covariant form which is preserved by a general spatial mapping. 
Very interestingly it has been shown that a very similar property 
does exist in other kinds of waves such as acoustic wave. Therefore we can have transformation acoustics which originates from the invariance of the acoustic wave equation under coordinate transformations \cite{chen2007acoustic}. In acoustics cloaking and more generally controlling acoustic waves propagation the medium parameters which must have spatial variations are pressure, mass density and bulk modulus.

Beside the classical waves with optical or acoustic natures, very recently it has been proven that quantum mechanical matter waves can also be controlled via spatial dependence of parameters like the effective masses and local potentials \cite{zhang2008cloaking}. This originates from covariance of Schr\"odinger equation with position dependent mass and potential terms governing the non-relativistic quantum mechanical waves. Therefore similar to electromagnetic and acoustic waves, we can achieve the matter waves cloaking by spatially controlling of the potentials and effective masses of the particles inside the cloak. 
Controlling matter waves in quantum mechanical systems has opened a new field which has been led to new phenomena and applications. For instance recently controlling of matter waves under the global Aharonov-Bohm effect and Cloaking spin matter waves have been investigated \cite{lin2009cloaking,lin2010cloaking}.

In this work, we find the coordinate transformation 
in order to have sphere, cylindrical, and ellipsoidal cloaks for electronic waves propagating according to the Schr\"odinger equation. Then using the inverse transformation the spatial form of effective mass tensor and the potential in a way that the wave propagation takes place equivalently to the corresponding deformed space in order to have the cloaking behavior. These findings could be useful in various applications including those related to electrical device engineerings based on semiconductors as well as quantum storages based on controllable cloaking by applying local gate voltages for instance \cite{esfar12,esfar13,alu13,alu13b,alu14,lee14}. In the remainder of the paper first we will introduce the theoretical formalism with more details. Then the results for the variety of exactly solvable situations with highly symmetric shapes will be presented. Finally the last section will be devoted to the conclusions. 

\section{Theoretical formalism}
We start the theoretical formulation of matter waves cloaking with the time independent Schr\"odinger equation with an effective mass tensor $\hat{m}$ which in general is position dependent \cite{aktacs2005exact,levy1995position,dekar1998exactly}. 
The time-independent Schr\"odinger equation for anisotropic materials is given by:
\begin{equation}
  -\frac{\hbar^2}{2}\nabla\cdot(\widehat{m}^{-1}\nabla\psi)+V\psi=E\psi
\label{e1}
\end{equation}
in which $V(\vec{r})$ indicates the potential profile.
In general effective mass tensor can be merged for instance due to the crystallographic structure and subsequently the general quadratic band dispersion for electrons inside the solids. In particular in certain semiconductors the effective masses has a tensorial form which can includes anisotropic off diagonal terms ($m_{ij}$). To obtain transformation equations, one can rewrite Eq. (\ref{e1}) as two coupled first order differential equations,
\begin{equation}
 \vec{u}=\hat{m}^{-1} \nabla \psi , \qquad - \frac{\hbar^2}{2}\nabla . \vec{u}=(E-V)\psi .
 \label{e2}
\end{equation}
In order to obtain the general covariant form of the Schr\"odinger equation in a curvilinear frame, we use the components denoted by $x^i$ and $x^{i^\prime}$ 
for the original and the transformed coordinate systems, respectively. For the sake of simplicity we
only focus on the orthogonal bases in which the metric tensor has a diagonal form. Writing down the divergence of the vector and the gradient of the wave function in the transformed coordinate, we obtain,
\begin{equation}
\big(- \frac{\hbar^2}{2} \big) \frac{1}{\sqrt{g}} \frac{\partial}{\partial x^{i^\prime}} \big(\frac{\sqrt{g}}{h_i m^{ij}h_j}\frac{\partial \psi}{ \partial x^{j^\prime}} \big) =(E-V)\psi ,
\label{e3}
\end{equation}
where $g=(h_1 h_2 h_3)^2=({\rm det}[h_{ij}])^2$ is the determinant of the metric tensor and $h_i=\frac{\partial \vec{x}}{\partial x^{i^\prime}}$ are the Lame coefficients $h_{ij}=h_i \delta _{ij}$ ($\delta _{ij}$ denotes the Kronecker delta). Because the Eq. (\ref{e3}) is covariant under coordinate transformation so its form will not change in any new coordinate system and subsequently we can write,
\begin{equation}
\big(- \frac{\hbar^2}{2} \big) \frac{\partial}{\partial x^{i^\prime}} \big(\frac{1}{m^{i^\prime j^\prime}}\frac{\partial \psi}{ \partial x^{j^\prime}} \big) =(E-V^\prime)\psi.
\label{e4}
\end{equation}
Comparing the two Eqs. (\ref{e3}) and (\ref{e4}) one can obtain the corresponding relation between the effective mass tensors and the scalar potentials in two different frames which are given by,
\begin{equation}
m^{i^\prime j^\prime}=\frac{h_i m^{ij} h_j}{\sqrt{g}} , \qquad V^\prime=E+\sqrt{g}(V-E) ,
\label{e5}
\end{equation}
\par
Now for any desired form of matter wave propagation it is sufficient to first find the suitable transformation from original Cartesian space to a deformed one. Then in application instead of deforming the space we can use the corresponding space dependent form of the mass tensor and the potentials. In particular using these equations, the general conditions for the matter wave cloaking can be obtained as we will discuss for certain geometries in the next section. Throughout the paper we focus on the elastic scatterings by the position dependent potentials and mass terms. Therefore we assume that the energy (frequency) of the quantum waves in the Eqs. (\ref{e5}) is constant.
\begin{figure}[htp]
\centering
\includegraphics[width=0.5\textwidth]{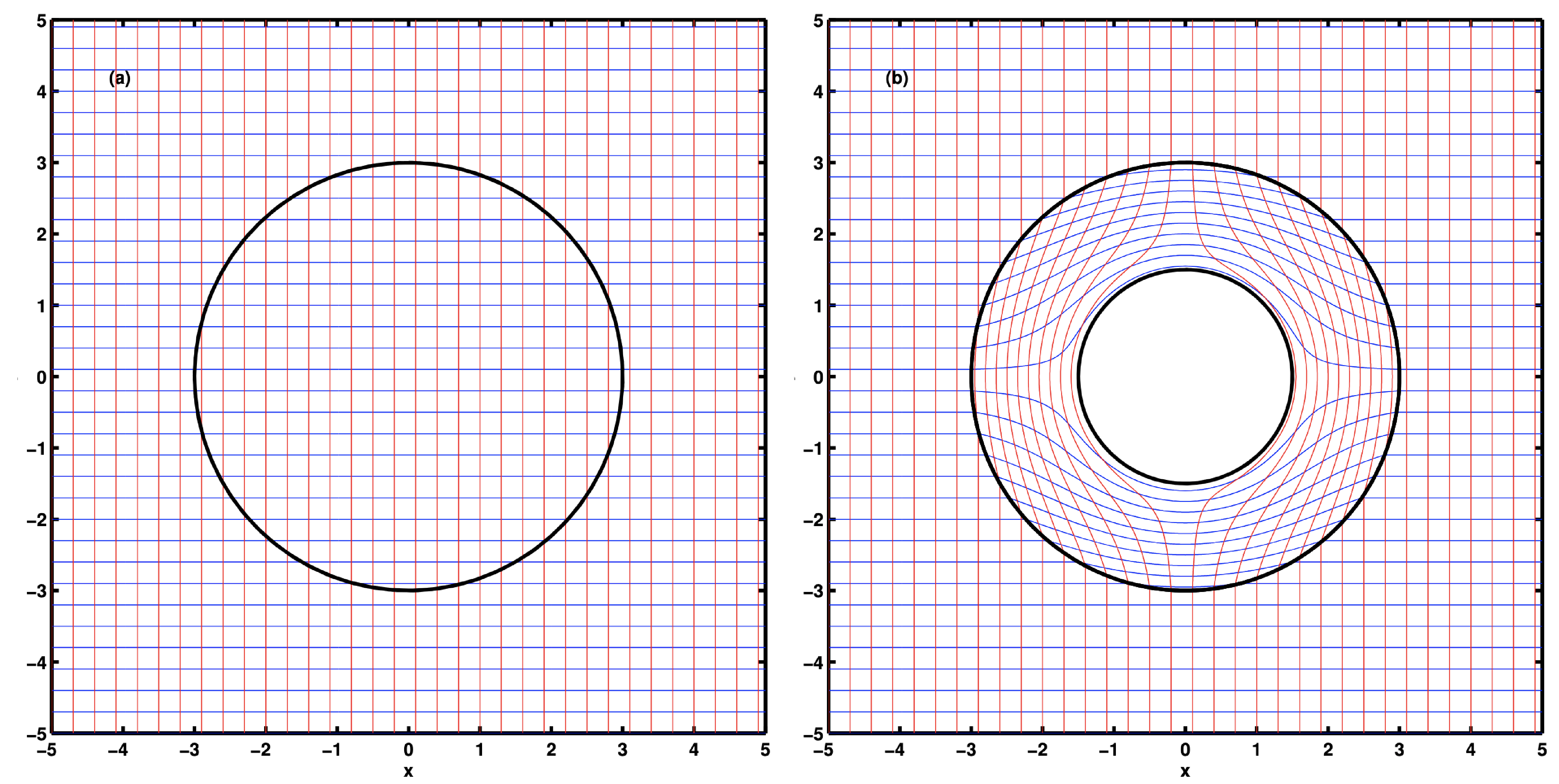}
\caption{(Color online) Application of coordinate transformation to a spherical region subjected to a
plane-wave electron propagating in the $x$-direction. As shown in the figures, the spatial transformation takes place in the $x-y$ plane. Figure (a) shows the closed region (area inside the circle) in Cartesian coordinates,
with vacuum region for both inside and outside regions of it. The rays of electron or generally matter waves (indicated by blue horizontal lines) and the wave fronts (shown by red perpendicular lines) of the incident propagating plane-wave are also shown. Figure (b) shows the transformed region according to
the transformation given by Eq. (\ref{e8}) with $b = 2a$ and the material properties in the cloaking region
determined by Eqs. (\ref{e13}).}
\label{Fig.1}
\end{figure}

\section{Results and discussion}
In the following we will present the exact analytical results for the coordinate transformations as well as the corresponding mass tensors and the potentials for a variety of matter wave cloaks with spherical, cylindrical, and ellipsoidal geometries. The clocking behavior is also illustrated using the shape of the cloaks and the behavior of propagating rays through them. 

\subsection{Spherical cloak}
The spherical cloak is the simplest cloak and so initially we consider a spherically symmetric transformation. This transformation compresses the space contained in volume of radius $b$ into a spherical shell of inner radius $a$ and outer radius $b$ and outside of the radius $b$ is not affected under the transformation (see Fig. \ref{Fig.1}). According to the transformation relations between $x^i$ and $x^{i^\prime}$ which indicate the old and new components, respectively, we find the coordinate transformation as below,
\begin{equation}
r^\prime=g(r) , \qquad \theta^\prime=\theta , \qquad \varphi^\prime=\varphi ,
\label{e6}
\end{equation}
where $g(r)$ is a monotonic radial scaling function. We note that when $r=a$ then $g(a)=0$ and also when $r=b$ then $g(r)=b$. Using the invariants of the transformation, the unit vectors in the original space and transformed space must be equal which means $\frac{x^\prime}{r^\prime}=\frac{x}{r}, \frac{y^\prime}{r^\prime}=\frac{y}{r}$ and $\frac{z^\prime}{r^\prime}=\frac{z}{r}$. these relations originates from the fact that  our transformation is radially symmetric and subsequently the following relation holds,
\begin{equation}
\frac{x^{i^\prime}}{r^\prime}=\frac{x^i}{r}\delta_i^{i^\prime},
\label{e7}
\end{equation}

Invoking the relationship between $r$ and $r^\prime$ [Eq.(\ref{e6})] we can write,
\begin{equation}
x^{i^\prime}=g(r)\frac{x^i}{r}\delta_i^{i^\prime},
\label{e8}
\end{equation}
Now using this equation, the transformation $\hat{h}$ is calculated with the following form,
\begin{align}
h_{ij} &= \delta_i^{i^\prime}\frac{\partial}{\partial x^j}\Big[g(r)\frac{x^i}{r}\Big] \nonumber \\
&= \Big[g^\prime(r)-\frac{g(r)}{r}\Big]\frac{x^ix^k}{r^2}\delta_{kj}\delta_i^{i^\prime}+\frac{g(r)}{r}\delta_i^{i^\prime},
\label{e9}
\end{align}
where $g^\prime(r)=\frac{\partial g(r)}{\partial r}$ and $r=(x^i x^k \delta_{ik})^\frac{1}{2}$. The components of the above expression can be also written  in the matrix form as,
\begin{equation}
\mathbf [h_{ij}]=\left(
\begin{array}{ccc}
f(r)\frac{x^2}{r^2}+\frac{g(r)}{r} &  f(r) \frac{xy}{r^2} & f(r)\frac{xz}{r^2} \\

f(r)\frac{yx}{r^2} & f(r)\frac{y^2}{r^2}+\frac{g(r)}{r} & f(r)\frac{yz}{r^2} \\

f(r)\frac{zx}{r^2} & f(r)\frac{zy}{r^2} & f(r)\frac{z^2}{r^2}+\frac{g(r)}{r}
\end{array} \right),
\label{e10}
\end{equation}
in which the function $f$ is defined as $f(r)=g^\prime(r)-\frac{g(r)}{r}$. Since the rotation cannot change the determinant of the matrix, therefore this property enables us to calculate the determinant using a rotation into a coordinate system such as $(x^i)=(r, 0, 0)$ in which all of the off-diagonal elements vanish. Then the determinant reads,
\begin{equation}
det[h_{ij}]=g^\prime(r) \Big[\frac{g(r)}{r}\Big]^2 .
\label{e11}
\end{equation}
\par
Having the transformation relation given above and using the Eq.(\ref{e5}), we can obtain the desired material properties for the spherical cloak. If our original space is free space then ${\hat{m}^*}{}^{ij}=m_o\delta^{ij}$ where $m_0$ is the mass in free space and $V=0$. Again, using a proper rotation, we are able to achieve the material properties as what follows,
\begin{align}
 \left(
\begin{array}{ccc}
m^{rr} & 0 & 0 \\
0 & m^{\theta \theta} & 0 \\
0 & 0 & m^{\varphi \varphi}
\end{array} \right)&=\frac{m_0}{g^\prime(r)} \Big[\frac{1}{g(r)}\Big]^2
 \left(
\begin{array}{ccc}
[rg^\prime(r)]^2 & 0 & 0 \\
0 & [g(r)]^2  & 0 \\
0& 0 & [g(r)]^2
\end{array} \right) ,\nonumber\\
V(r,E)&=\Big[ 1- g^\prime(r) \Big( \frac{g(r)}{r} \Big)^2 \Big]E .
\label{e12}
\end{align}
In order to obtain the final result assuming a linear scaling function such as $g(r)=\frac{b-a}{b}r+a$ which is shown in Fig. \ref{Fig.1}), the mass tensor and the potentials as functions of positions are obtained with the following forms,
\begin{align}
m^{rr} &= \frac{bm_o}{b-a}\Big(\frac{r^\prime-a}{r^\prime}\Big)^2,\nonumber \\
m^{\theta \theta} &=\frac{bm_o}{b-a},\nonumber  \\
m^{\varphi \varphi} &=\frac{bm_o}{b-a},\nonumber \\
V(r,E)&=\Big[1-\Big(\frac{b-a}{b}\Big)^3\Big(\frac{r^\prime}{r^\prime-a}\Big)^2\Big]E.
\label{e13}
\end{align}

Figure \ref{Fig.1}(a) indicates the Cartesian grid and spherical region which is mapped to the curved grid of
physical space shown in the Fig. \ref{Fig.1}(b). The physical coordinates enclose a spherical hole that is made for cloaking. Consequently, a medium that performs this transformation acts as an invisibility device. The region with radius $a$ is invisible and  $b-a$ describes the thickness of the cloaking layer. A coordinate transformation Eq.(\ref{e8}) with $b=2a$ transforms this closed region into the cloaking region shown in Fig. \ref{Fig.1}(b). The grid lines in this region transform accordingly, and the transformed grid lines which are geodesics in the transformed coordinates, can be interpreted as matter wave rays and wave fronts in the presence of the cloak. As expected, the matter waves never penetrate into the cloaked region. Evidently, the matter wave rays and wave fronts beyond the outer boundary remain the same, since an identity transformation is applied in this region.

\subsection{Cylindrical cloak }
The matter wave cloaking for  the cylindrical geometry and transformation relations is the same as the spherical case. Since the transformation in the $x-y$ plane is the same as the spherical case and the transformation in the $z$ direction remains the identity, so the transformation matrix can be written as,
\begin{equation}
\mathbf[h_{ij}]=\left(
\begin{array}{ccc}
[s^\prime(\rho)-\frac{s(\rho)}{\rho}]\frac{x^2}{\rho^2}+\frac{s(\rho)}{\rho} &  [s^\prime(\rho)-\frac{s(\rho)}{\rho}] \frac{xy}{\rho^2} & 0 \\

[s^\prime(\rho)-\frac{s(\rho)}{\rho}]\frac{yx}{\rho^2} &[s^\prime(\rho)-\frac{s(\rho)}{\rho}]\frac{y^2}{\rho^2}+\frac{s(\rho)}{\rho} & 0 \\

0 & 0 & 1
\end{array} \right).
\label{e14}
\end{equation}
Here $s(\rho)$ is a monotonic radial scaling function and relation between the two coordinates is given by,
\begin{equation}
\rho^\prime=s(\rho) , \qquad \varphi ^\prime=\varphi , \qquad z^\prime=z.
\label{e15}
\end{equation}
Again by rotating into a system where the off-diagonal elements vanish, $(x^i)=(\rho , 0 , 0)$, we can easily calculate the determinant for the cylindrical coordinate,
\begin{equation}
det[h_{ij}]=\frac{s^\prime(\rho)s(\rho)}{\rho} .
\label{e16}
\end{equation}
Next using the Eq.(\ref{e5}), we can obtain the spatial variations of the material properties in order to behave as a cylindrical cloak,
\begin{align}
 \left(
\begin{array}{ccc}
m^{\rho \rho} & 0 & 0 \\
0 & m^{\varphi \varphi} & 0 \\
0 & 0 & m^{zz}
\end{array} \right)&=\frac{\rho m_0}{s^\prime(\rho) s(\rho)}
 \left(
\begin{array}{ccc}
[s^\prime(\rho)]^2 & 0 & 0 \\
0 & [\frac{s(\rho)}{\rho}]^2  & 0 \\
0& 0 & 1
\end{array} \right) ,\nonumber\\
V(r,E)&=\Big[1- \frac{s^\prime(\rho) s(\rho)}{\rho} \Big]E ,
\label{e17}
\end{align}
Considering a linear scaling function such as $s(\rho)=\frac{b-a}{b}\rho+a$, the mass and potential profile for the cylindrical cloak is obtained with the following relations,
\begin{align}
m^{\rho \rho}&=\frac{\rho^\prime-a}{\rho^\prime}m_0 ,\nonumber \\
m^{\varphi \varphi}&=\frac{\rho^\prime}{\rho^\prime-a}m_0 , \nonumber\\
m^{zz}&=\Big(\frac{b}{b-a}\Big)^2 \Big(\frac{\rho^\prime-a}{\rho^\prime}\Big)m_0 ,\nonumber \\
V(\rho,E)&=\Big[ 1- \Big(\frac{b-a}{b}\Big)^2\Big( \frac{\rho^\prime}{\rho^\prime-a}\Big)\Big]E.
\label{e18}
\end{align}

\subsection{Ellipsoidal cloak}
\begin{figure}
\centering
\includegraphics[width=0.99\textwidth]{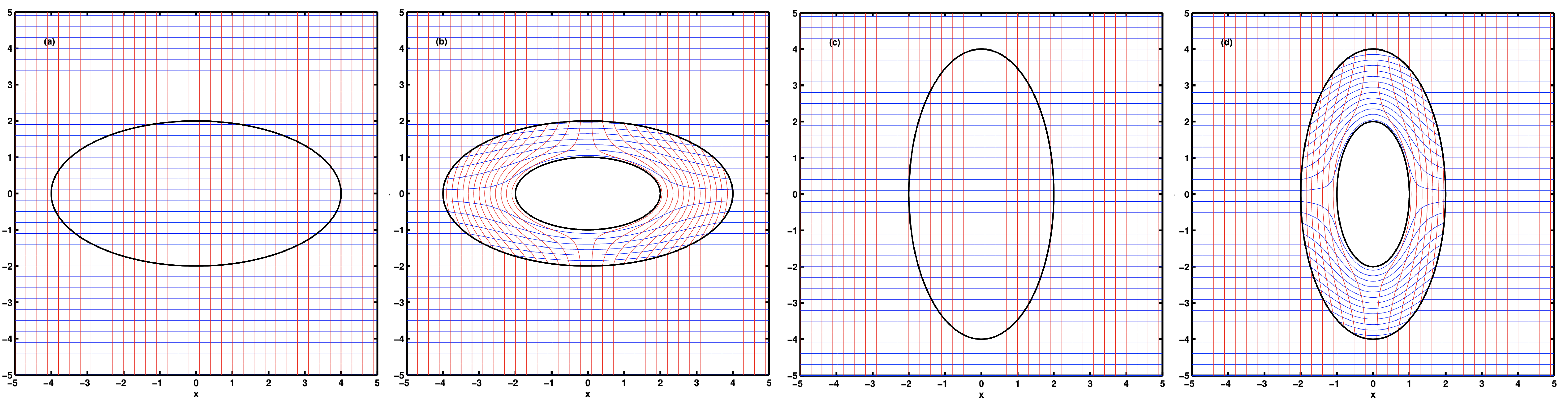}
\caption{(Color online) Similar to Fig. \ref{Fig.1} but for the ellipsoidal cloaks. Figure (a) and (c) show the closed region (area inside the horizontal and vertical ellipsoids) in Cartesian coordinates. Figures (b) and (d) show the transformed region according to
the transformation Eq. (\ref{e20}) with $b = 2a$ and the material properties in the cloaking region
determined by Eqs. (\ref{e24}). The matter wave rays and wave fronts transform accordingly.}
\label{Fig.2}
\end{figure}

The invisibility cloak for quantum mechanical waves in the ellipsoidal geometry is defined as an ellipsoidal shell that any point on the outer and inner boundaries of it have to justify the following scaled distance,
\begin{equation}
r=\Bigg(\sum_{i=1}^3 \Big(\frac{x^i}{\alpha_i}\Big)^2 \Bigg)^\frac{1}{2} ,
\label{e19}
\end{equation}
where $\alpha_1 , \alpha_2$ and $\alpha_3$ are positive numbers describing the aspect ratio of the ellipsoid. Eq. (\ref{e19}) guarantees $r=a$ for any point $x^i$ on the inner boundary and $r=b$ for any point $x^i$ on the outer boundary $(a<b)$. The transformation matrix in this case can be readily obtained using a generalized form of Eq.(\ref{e8}) for the ellipsoids. Therefore, the coordinate transformation reads,
\begin{equation}
x^{i^\prime}=\frac{r^\prime}{r}x^i \delta_i^{i^\prime} .
\label{e20}
\end{equation}
Here $r^\prime=\frac{b-a}{b}r+a$ and $r=\Big(\sum_{i=1}^3 \big(\frac{x^i}{\alpha_i}\big)^2 \Big)^\frac{1}{2}$ as seen in Fig. \ref{Fig.2}. These relations lead to the following transformation matrix elements,
\begin{equation}
h_{ij}=\frac{r^\prime}{r} - \frac{ax^ix^k}{\alpha_j^2 r^3}\delta_{kj}\delta_i^{i^\prime} ,
\label{e21}
\end{equation}
where $\frac{\partial}{\partial x^j}(r^{-1})=-\frac{1}{r^3}\frac{x^j}{\alpha_j^2}$. Finally, the transformation matrix is,
\begin{equation}
\mathbf[h_{ij}]=\left(
\begin{array}{ccc}
\frac{r^\prime}{r}-\frac{ax^2}{\alpha_1^2 r^3} & -\frac{axy}{\alpha_2 r^3} & -\frac{axz}{\alpha_3^2 r^3} \\

-\frac{ayx}{\alpha_1^2 r^3 }& \frac{r^\prime}{r}-\frac{ay^2}{\alpha_2^2 r^3} & -\frac{ayz}{\alpha_3^2 r^3} \\

-\frac{azx}{\alpha_1^2 r^3} & -\frac{azy}{\alpha_2^2 r^3} & \frac{r^\prime}{r}-\frac{az^2}{\alpha_3^2 r^3}
\end{array} \right).
\label{e22}
\end{equation}
By rotating into a new system where the off-diagonal elements vanish, \textit{i.e.} $(x^i)=(\alpha_1r, 0, 0)$, we can easily find the determinant,
\begin{equation}
det[h_{ij}]=\Big( \frac{r^\prime}{r}\Big)^2 \frac{b-a}{b}.
\label{e23}
\end{equation}

Now using the transformation (\ref{e22}) and invoking the relations (\ref{e5}) we can find the spatial form of the mass tensor elements and the potential profile in order to have ellipsoidal cloaking, 
\begin{align}
m^{x' x'}&=\frac{bm_0}{b-a}\big[ 1+\frac{a^2 r'{}^2-2a r'{}^3}{\alpha_1^2 r'^6}x'^2 \big] ,\nonumber\\
m^{x' y'}&=\frac{bm_0}{b-a}\big[\frac{a^2 r'{}^2-2a r'{}^3}{\alpha_2^2 r'^6}x' y'  \big] ,\nonumber\\
m^{x' z'}&=\frac{bm_0}{b-a}\big[\frac{a^2 r'{}^2-2a r'{}^3}{\alpha_3^2 r'^6}x' z'  \big] ,\nonumber\\
m^{y' x'}&=\frac{bm_0}{b-a}\big[\frac{a^2 r'{}^2-2a r'{}^3}{\alpha_1^2 r'^6}y' x'  \big] ,\nonumber\\
m^{y' y'}&=\frac{bm_0}{b-a}\big[ 1+\frac{a^2 r'{}^2-2a r'{} ^3}{\alpha_2^2 r'^6}y'^2 \big] ,\nonumber\\
m^{y' z'}&=\frac{bm_0}{b-a}\big[\frac{a^2 r'{}^2-2a r'{}^3}{\alpha_3^2 r'^6}y' z'  \big] ,\nonumber\\
m^{z' x'}&=\frac{bm_0}{b-a}\big[\frac{a^2 r'{}^2-2a r'{}^3}{\alpha_1^2 r'^6}z' x'  \big] ,\nonumber\\
m^{z' y'}&=\frac{bm_0}{b-a}\big[\frac{a^2 r'{}^2-2a r'{}^3}{\alpha_2^2 r'^6}z' y'  \big] ,\nonumber\\
m^{z' z'}&=\frac{bm_0}{b-a}\big[ 1+\frac{a^2 r'^2-2a r'^3}{\alpha_3^2 r'{}^6}z'{}^2 \big] ,\nonumber\\
V(r,E)&=\Big[1-\Big(\frac{b-a}{b}\Big)^3 \Big( \frac{r'}{r'-a} \Big)^2\Big]E.
\label{e24}
\end{align}
When $\alpha_1=\alpha_2=\alpha_3=1$, the ellipsoidal cloak reduces to a spherical cloak. The similar results for a matter wave cloaking with elliptical geometry in two dimensions can also be easily found.
Figure \ref{Fig.2}(a) shows the Cartesian grid and spheroidal region that is mapped to the curved grid of
physical space shown in the Fig. \ref{Fig.2}(b). The physical coordinates enclose a elliptical hole which is made for cloaking. A coordinate transformation Eq.(\ref{e20}) with $b=2a$ transforms this closed region into the cloaking region shown in Fig. \ref{Fig.2}(b). Evidently, the matter wave rays and wave fronts beyond the outer boundary remain the same when an identity transformation is applied in this region. In Fig. \ref{Fig.2}(c) and (d), the similar results have been obtained for vertical ellipsoidal cloak.

\section{Conclusions}
We have studied invisibility cloaks for matter waves with three geometries using the coordinate transformation approach. The effective mass tensor and macroscopic potential have been obtained for spherical, cylindrical and ellipsoidal cloaks. Previously people have found similar results for the permittivity and permeability tensors and a comparison between their works and ours shows that the generalization of the results for the quantum mechanical waves along with classical electromagnetic and acoustic ones. Notably the scalar potential for the spherical and ellipsoidal cloaks has the same form. The behavior of matter waves has been demonstrated for several specific geometries using the coordinate transformation for the cloaks and the rays of propagating matter waves. 

\bibliography{mybibfile} 

%

\end{document}